\documentclass[3p,twocolumn]{elsarticle}

\usepackage{amsfonts,amsmath,amssymb}
\usepackage{graphicx,color}

\def\blfootnote{\xdef\@thefnmark{}\@footnotetext}
\newcommand{\ie}{\textit{i.e.}}

\begin{document}

\begin{frontmatter}

%\title{A Resolution to the Charm-Baryon Puzzle in Proton-Proton Collisions?}
\title{Charm-Baryon Production in Proton-Proton Collisions}

\author[a]{Min He}
\author[b]{Ralf Rapp}

\address[a]{Department of Applied Physics, Nanjing University of Science and Technology, Nanjing 210094, China}
\address[b]{Cyclotron Institute and Department of Physics \& Astronomy,
Texas A\&M University, College Station, TX 77843, USA}

\date{\today}

\begin{abstract}
Recent measurements of charm-baryon production in proton-proton collisions at the LHC have found a surprisingly large yield relative to those of $D$-mesons. We propose that this observation can be explained by the statistical hadronization model (SHM), by employing a largely augmented set of charm-baryon states beyond the current listings of the particle data group. We estimate the additional states using guidance from the relativistic quark model and from lattice QCD. Using charm- and strange-quark fugacity factors to account for the well-known suppression of heavy flavor in elementary collisions, we compute the yields and spectra of $D$, $D_s$ and $\Lambda_c$ hadrons in proton-proton collisions at $\sqrt{s}=5$\,TeV. Our main finding is that the enhanced feeddown from excited charm baryons can account for the $\Lambda_c/D^0$ ratio measured by ALICE at midrapidity, with some caveat for the forward-rapidity LHCb data. Furthermore, assuming independent fragmentation of charm quarks but with the hadronic ratios fixed by the SHM, the measured transverse-momentum ($p_T$) spectra of $D$-mesons and $\Lambda_c$ can also be described; in particular, the low-$p_T$ enhancement in the observed $\Lambda_c/D^0$ ratio is attributed to the enhanced feeddown from ``missing" charm-baryon states. We comment on the implications of these findings for measurements of $D_s$ and $\Lambda_c$ in heavy-ion collisions.
\end{abstract}


\end{frontmatter}

%%%%%%%%%%%%%%%%%%%%%%%%%%%%%%%%%%%%%%%%%%%%%%%%%%%%%%%%
\section{Introduction
\label{sec_intro}}
%%%%%%%%%%%%%%%%%%%%%%%%%%%%%%%%%%%%%%%%%%%%%%%%%%%%%%%%

The production of heavy-flavor (HF) particles in high-energy hadronic collisions is a versatile source of information on various aspects of Quantum Chromodynamics (QCD). The primordial pair production of heavy quarks and their anti-quarks in a hard partonic scattering event is a fruitful testing ground for perturbative QCD calculation. Based on collinear factorization, this process essentially governs the total heavy-quark (HQ) production cross section. On the other hand, the subsequent hadronization of charm ($c$) and bottom ($b$) quarks into HF hadrons is an inherently nonperturbative process related to, or even driven by, the confining property of QCD.

Various models to account for the different species of the observed HF hadrons have been put forward, including independent fragmentation models, usually applicable at sufficiently large momentum, or color neutralization models such as string fragmentation, color reconnection or color ropes~\cite{Cacciari:1998it,Kniehl:2005mk,Frixione:2007nw,Kniehl:2012ti,Cacciari:2012ny,Kramer:2017gct,Maciula:2018iuh,Helenius:2018uul}. In addition, the statistical hadronization model (SHM) has been employed, essentially
replacing the complexity of the hadronization process by thermo-statistical weights governed by the masses of available hadron states
at a universal hadronization ``temperature", $T_H$. The SHM has been successfully applied to light- and strange-hadron production in both heavy-ion and elementary collisions, with the addition of a strangeness suppresion factor, $\gamma_s<1$, in the latter (and in peripheral heavy-ion collisions)~\cite{BraunMunzinger:2003zd,Andronic:2017pug}. It also works for various charm-meson ratios~\cite{Andronic:2007zu,Andronic:2009sv,Acharya:2018hre,Adam:2016ich,Acharya:2017jgo,Acharya:2019mgn}.
However, recent mesurements of charm-baryon production in proton-proton ($pp$) collisions by the ALICE collaboration at the LHC held a surprise; specifically, the cross section ratio of prompt  $\Lambda_c$ over $D^0$-mesons, $\Lambda_c/D^0\simeq0.54$, measured at
$\sqrt{s}=7$\,GeV~\cite{Acharya:2017kfy} turns out to be much
larger than expected in most event generators~\cite{Kniehl:2005mk,Frixione:2007nw,Kniehl:2012ti,Christiansen:2015yqa,Bierlich:2015rha}, as well as in the
SHM where it is $\sim0.22$~\cite{Andronic:2007zu} based on charm-hadron states listed by the particle data group (PDG)~\cite{Tanabashi:2018oca}.
Recent attempts~\cite{Maciula:2018iuh} to reproduce these data
using an independent fragmentation approach confirmed the challenge to describe the ALICE data. Measurements of $\Lambda_c$ production have also been carried out by the LHCb
collaboration at forward rapidities, yielding smaller values of $\Lambda_c/D^0\simeq0.25\pm0.05$ in $\sqrt{s}=7$\,TeV $pp$ collisions~\cite{Aaij:2013mga}, and $\sim0.35\pm0.05$ in $\sqrt{s_{\rm NN}}=5$\,TeV $p$Pb collisions~\cite{Aaij:2018iyy}.

In the present paper we explore in how far the observed enhancement of $\Lambda_c$ production can be due to hitherto unobserved charm-baryon states, not listed in the PDG tables~\cite{Tanabashi:2018oca}. For example, the latter currently contain 6 $\Lambda_c$ and 3 $\Sigma_c$ states, compared to 14 $\Lambda$ and 10 $\Sigma$ states (plus additional less certain states)
in the strangeness sector.
All of the observed excited single-charm baryons have dominant decay branchings into $\Lambda_c$ final states with widths of the order of MeV, and thus their ``feeddown" qualifies as ``prompt" $\Lambda_c$ production as measured in experiment.
We will estimate the missing states by taking guidance from relativistic quark model (RQM)~\cite{Ebert:2011kk} calculations. We will
implement the updated thermal yields to compute the hadro-chemistry of $\Lambda_c$, $D^0$, $D^*$ and $D_s$ yields, and also calculate their transverse-momentum ($p_T$) spectra through fragmentation functions of a given $c$-quark spectrum adapted for
mesons and baryons in the fixed-order-next-to-leading-log (FONLL) scheme~\cite{Cacciari:1998it,Cacciari:2012ny}.

%%%%%%%%%%%%%%%%%%%%%%%%%%%%%%%%%%%%%%%%%%%%%%%%%%%%%%%%%%%%%%%%
\section{Charm-Baryon Spectrum and SHM
\label{sec_shm}}
%%%%%%%%%%%%%%%%%%%%%%%%%%%%%%%%%%%%%%%%%%%%%%%%%%%%%%%%%%%%%%%%
The issue of ``missing resonances" is a long-standing problem in hadronic spectroscopy, in particular in the baryon sector~\cite{Crede:2013sze}. For charm baryons, this problem is particularly challenging as direct spectroscopic measurements are rather scarce. Lattice-QCD (lQCD) computations~\cite{Padmanath:2014bxa} of the charm-baryon spectrum indeed show a vastly richer spectrum than currently measurable, with ten's of additional states in the single-charm sector (most pertinent to our present work), approximately following quark model classifications of SU(6)-O(3) flavor-spin-angular-momentum symmetry. More indirectly, the analysis of the partial pressure of open-charm states and charm-quark susceptibilities in thermal lQCD~\cite{Bazavov:2014yba} also found that their results
for temperatures $T$=150-170\,MeV are much under-predicted using PDG states only, while the use of a charm-hadron spectrum
predicted by the RQM~\cite{Ebert:2011kk} resulted in a good description.

Motivated by these findings we construct a SHM based on two different inputs for the charm-hadron states: (a) a PDG version
of only including states listed in
Ref.~\cite{Tanabashi:2018oca}, and (b) a RQM version including additionally predicted charm-baryon states as listed
in Ref.~\cite{Ebert:2011kk}, which amounts to an extra 18 $\Lambda_c$'s, 42 $\Sigma_c$'s, 62 $\Xi_c$'s, and 34 $\Omega_c$'s up to a mass of $3.5$\,GeV. We have checked that including additional RQM mesons would increase the thermal density of $D^0$ by $\sim 10\%$, which does not affect our final results significantly. On the other hand, the baryon states in the RQM  are based on a light-diquark scheme, which tends to give fewer states than a genuine three-quark picture~\cite{Ebert:2011kk} which could counter-balance an increase in excited $D$-meson states. As usual in the SHM, the thermal hadron densities follow from their masses, $m_i$, and
spin-isospin degeneracies, $d_i$, evaluated at a hadronization temperature, $T_{H}$, as
\begin{equation}
\label{thermal_density}
n_i=\frac{d_i}{2\pi^2}m_i^2T_{H}K_2(\frac{m_i}{T_H}) \ ,
\end{equation}
where $K_2$ is the modified Bessel function of second order. Given the agreement of the lQCD susceptibilities with the same RQM charm-baryon ensemble as used here up to temperatures of 170\,MeV~\cite{Bazavov:2014yba}, we use the latter as an upper estimate of $T_H$, and utilize lower values as part of our error estimate. A flavor hierarchy in the operational hadronization temperature of the
QCD crossover transition has been suggested before based on comparisons of light and strange-quark susceptibilites~\cite{Bellwied:2013cta}, amounting to an upward shift of about 15\,MeV for strange hadrons.
%kthe contribution of generally more massive baryons to the pressure more pronounced than using a lower, {\it e.g.} 160\,MeV closer to the light flavor chiral transition pseudo-docritical temperature.

An important ingredient are the branching ratios (BRs) of the excited charm hadrons to their ground states. For observed states, we use BRs as available from the PDG, and for ``seen" decay channels without BRs we assume an equal weight.
To explore the maximum effect of all RQM $\Lambda_c$'s and $\Sigma_c$'s not listed by the PDG, we assume that their decay chain always ends up with a ground-state $\Lambda_c^+$ (plus one or two $\pi$'s). This is motivated by the fact that the ground-state $\Sigma_c(2455)$ is listed with a $\sim$100\% BR into $\Lambda_c+\pi$, and by chiral-quark model studies~\cite{Zhong:2007gp} where the BR of $D+N$ channels for several studied highly excited $\Lambda_c$'s and $\Sigma_c$'s were predicted to be very small compared to $\Lambda_c + n\pi$ channels. For excited $\Xi_c$'s (containing one strange quark) the PDG indicates $\Lambda_c + K$ decay channels; lacking quantitative knowledge of those, we assume a 50\% BR for the additional RQM $\Xi_c$'s decaying to $\Lambda_c$, with the remaining 50\% to the ground state $\Xi_c$. Finally, for the thermal densities of both $D_s$ mesons and $\Xi_c$ baryons, containing one strange (anti)quark, we apply a strangeness suppression factor of $\gamma_s=0.6$ in Eq.~(\ref{thermal_density}), in line with the empirical value of $0.56\pm 0.04$ extracted from $\sqrt{s}=200$\,GeV $pp$ collisions~\cite{Abelev:2008ab} (for $\Omega_c$'s, $\gamma_s^2$ is applied accordingly).

\begin{table*}[!t]
\begin{center}
\begin{tabular}{lccccccc}
\hline\noalign{\smallskip}
% $~~~$
$n_i$~($\cdot 10^{-4}~\rm fm^{-3}$)          &$D^0$  &$D^+$ &$D^{*+}$  &$D_s^+$ &$\Lambda_c^+$  &$\Xi_c^{+,0}$  &$\Omega_c^0$\\
\noalign{\smallskip}\hline\noalign{\smallskip}
PDG(170)  & 1.161   & 0.5098  &   0.5010   &  0.3165  &  0.3310  & 0.0874 & 0.0064  \\
PDG(160)  & 0.4996  & 0.2223  &   0.2113   &  0.1311  &  0.1201  & 0.0304 & 0.0021  \\
RQM(170)  & 1.161   & 0.5098  &   0.5010   &  0.3165  &  0.6613  & 0.1173 & 0.0144 \\
RQM(160)  & 0.4996  & 0.2223  &   0.2113   &  0.1311  &  0.2203  & 0.0391 & 0.0044 \\
\noalign{\smallskip}
\hline
\end{tabular}
\end{center}
\caption{Thermal densities of ``prompt" ground-state charmed hadrons for hadronization temperatures of $T_{H}$=170 and 160\,MeV (including strong feeddowns) in the PDG and RQM scenarios.}
\label{tab_dens}
\end{table*}
The calculated thermal densities (with strong feeddowns) of the ground-state charm hadrons are summarized in Tab.~\ref{tab_dens},
where we also include results for $T_{H}=160$\,MeV.
The densities are converted into fractions of the total charm content in Tab.~\ref{tab_frac}.
The additional baryon states in the RQM much enhance the fraction of the ground-state $\Lambda_c$ in the system, relative to the
PDG scenario, by about $\sim 73\%~(65\%)$ at $T_{H}=170~(160)$\,MeV. We furthermore compute the ratios of $D^+$, $D^{*+}$,
$D_s^+$ and $\Lambda_c^+$ to the $D^0$, as summarized in Tab.~\ref{tab_rat}. The meson ratios are rather stable with respect to temperature variations, but the baryon-to-meson ratio is more sensitive. In the PDG scenario with $T_{H}=160$\,MeV, $\Lambda_c^+/D^0\simeq0.24$, close to the previously reported SHM value of 0.22 obtained for $T_{H}$=156.5\,MeV~\cite{Andronic:2007zu,Andronic:2018}. This value is increased to $\Lambda_c^+/D^0\simeq0.57$ at $T_{H}=170$\,MeV in the RQM scenario, almost doubling the PDG value and becoming comparable to the ALICE measurement~\cite{Acharya:2017kfy}.
This is one of the main results of our work.
\begin{table}[!t]
\label{tab_frac}
\begin{tabular}{lcccccc}
\hline\noalign{\smallskip}
% $~~~$
$f_i$                      &$D^0$  &$D^+$ & $D_s^+$ & $\Lambda_c^+$ \\
\noalign{\smallskip}\hline\noalign{\smallskip}
PDG(170)          & 0.4813 & 0.2113  &  0.1312  &  0.1372 \\
PDG(160)          & 0.4968 & 0.2210  &  0.1304  &  0.1194 \\
RQM(170)          & 0.4175 & 0.1834  &  0.1138  &  0.2379 \\
RQM(160)          & 0.4473 & 0.1990  &  0.1174  &  0.1973 \\
\noalign{\smallskip}\hline
\end{tabular}
\caption{Thermal fractions of $D^0$, $D^+$, $D_s^+$ and $\Lambda_c^+$ for hadronization temperatrues of $T_{H}$=170 and 160\,MeV (including strong feeddowns) in the PDG and RQM scenarios.}
\end{table}

In the following, we will keep the RQM scenario with $T_{H}=170$\,MeV as our default and calculate the $p_T$-differential cross sections of charmed hadrons by fragmenting a universal underlying charm-quark $p_t$ spectrum.

\begin{table}[!t]
\begin{tabular}{lcccccc}
\hline\noalign{\smallskip}
% $~~~$
$r_i$        &$D^+/D^0$    &$D^{*+}/D^0$    &$D_s^+/D^0$   &$\Lambda_c^+/D^0$ \\
\noalign{\smallskip}\hline\noalign{\smallskip}
PDG(170)      & 0.4391  & 0.4315  & 0.2736   & 0.2851 \\
PDG(160)      & 0.4450  & 0.4229  & 0.2624   & 0.2404 \\
RQM(170)      & 0.4391  & 0.4315  & 0.2726   & 0.5696 \\
RQM(160)      & 0.4450  & 0.4229  & 0.2624   & 0.4409 \\

\noalign{\smallskip}\hline
\end{tabular}
\caption{Ratios of $D^+$, $D^{*+}$, $D_s^+$ and $\Lambda_c^+$ to $D^0$ at $T_{H}$=170 and 160\,MeV (including strong feeddowns) in the PDG and RQM scenarios at two different hadronization temperatures.}
\label{tab_rat}
\end{table}

%%%%%%%%%%%%%%%%%%%%%%%%%%%%%%%%%%%%%%%%%%%%%%%%%%%%%%%%%%%%%%%%
\section{Fragmenation and Decay Simulation
\label{sec_frag}}
%%%%%%%%%%%%%%%%%%%%%%%%%%%%%%%%%%%%%%%%%%%%%%%%%%%%%%%%%%%%%%%%
As discussed in the introduction, the charm pair production, as a hard process, is believed to be governed by perturbative QCD,
even down to low momenta. Therefore, our starting point to compute charm-hadron $p_T$ spectra is the charm-quark $p_t$ spectrum in $pp$ collisions at $\sqrt{s}=5.5$\,TeV as simulated by FONLL~\cite{Cacciari:1998it,Cacciari:2012ny} (we use it as a proxy for that at $\sqrt{s}=5.02$\,TeV).
We utilize it to perform fragmentation  into various charmed
mesons and baryons using the fragmentation
function~\cite{Braaten:1994bz} that was also implemented
in the FONLL framework,
\begin{align}
\label{frag}
D_{c\rightarrow H}(z)=N\frac{rz(1-z)^2}{[1-(1-r)z]^6}[(6-18(1-2r)z
 \nonumber\\
+(21-74r+68r^2)z^2 \qquad \quad
\nonumber\\
-2(1-r)(6-19r+18r^2)z^3
\nonumber\\
\quad+3(1-r)^2(1-2r+2r^2)z^4] \ ,
\end{align}
where $z=p_T/p_t$ is the fraction of the hadron ($H$) momentum ($p_T$) relative to the quark momentum ($p_t$), and the
parameter $r$ may be interpreted as the ratio of the mass of the fragmenting quark to the mass of the
hadron~\cite{Braaten:1994bz}. The normalization, $N$, of the fragmentation function into each hadron is, however, determined according to the pertinent thermal densities calculated in the
RQM scenario as described in the previous section. The assumption here is that the phase-space population implied by the thermal model does not significantly affect the $p_T$ spectrum of the hadrons emerging from the fragmentation of the primordial  charm-quark spectrum (some of that effect is absorbed into our tuning of the $r$ parameter described below).

We tune the parameter $r$ in Eq.~(\ref{frag}) for the ground-state $D^0$ and $\Lambda_c^+$ as to fit the experimental slope of their $p_T$ spectra. Once $r_{D^0}$ is fixed, the value of $r$ for other $D$ and $D_s$ mesons ($M$) follows from mass scaling: $r_M/r_{D^0}=((m_M-m_c)/m_M)/((m_{D^0}-m_c)/m_{D^0})$~\cite{Braaten:1994bz}, where $m_c=1.5$\,GeV is the charm-quark mass used in our calculations. The same is done for charm baryons ($B$) based on $r_{\Lambda_c^+}$: $r_B/r_{\Lambda_c^+}=((m_B-m_c)/m_B)/((m_{\Lambda_c^+}-m_c)/m_{\Lambda_c^+})$. Through our fits the best $r$-values for the ground-state hadrons turn out to be
$r_{D^0}$=0.1 and $r_{\Lambda_c^+}$=0.16. Each charm hadron formed from fragmentation is then decayed into
ground-state particles assuming a constant matrix element, with the decay kinematics solely determined by phase space,
and the pertinent branching ratios discussed in Sec.~\ref{sec_shm}.

To effectively conduct the fragmentation and decay simulations, we introduce an ``average" baryon state to represent the
additional RQM states of each category (\ie, with the same isospin) by  a combined spin degeneracy as the sum of the pertinent
category, and an average mass that results in a thermal density corresponding to the sum of all states in that category.
Specifically, 18 additional $\Lambda_c$'s are represented by an ``average" $\bar\Lambda_c^{*}$ of effective mass
$3.17$\,GeV and total spin degeneracy of 43.5; 42 additional $\Sigma_c$'s by an ``average" $\bar\Sigma_c^{*}$ of
effective mass $3.10$\,GeV and total spin degeneracy 88.5;  62 additional $\Xi_c$'s  by an ``average" $\bar\Xi_c^{*}$ of
effective mass $3.24$\,GeV and total spin degeneracy 135.5; and additional 34 $\Omega_c$'s  by an ``average"
$\bar\Omega_c^{*}$ of effective mass $3.26 $\,GeV and total spin degeneracy 65.5.
To check the accuracy of this mass-averaging procedure, we have calculated the integrated yields of each
ground-state particle from fragmentation plus decay simulations, and confirmed that the pertinent fractions and
ratios agree with those calculated from the explicit RQM particle content (as listed in Tabs.~\ref{tab_frac} and \ref{tab_rat})
within a few percent.

%%%%%%%%%%%%%%%%%%%%%%%
\subsection{LHC     \label{ssec_lhc}}
%%%%%%%%%%%%%%%%%%%%%%%
%
\begin{figure} [!t]

\vspace{-0.6cm}

\includegraphics[width=1.0\columnwidth]{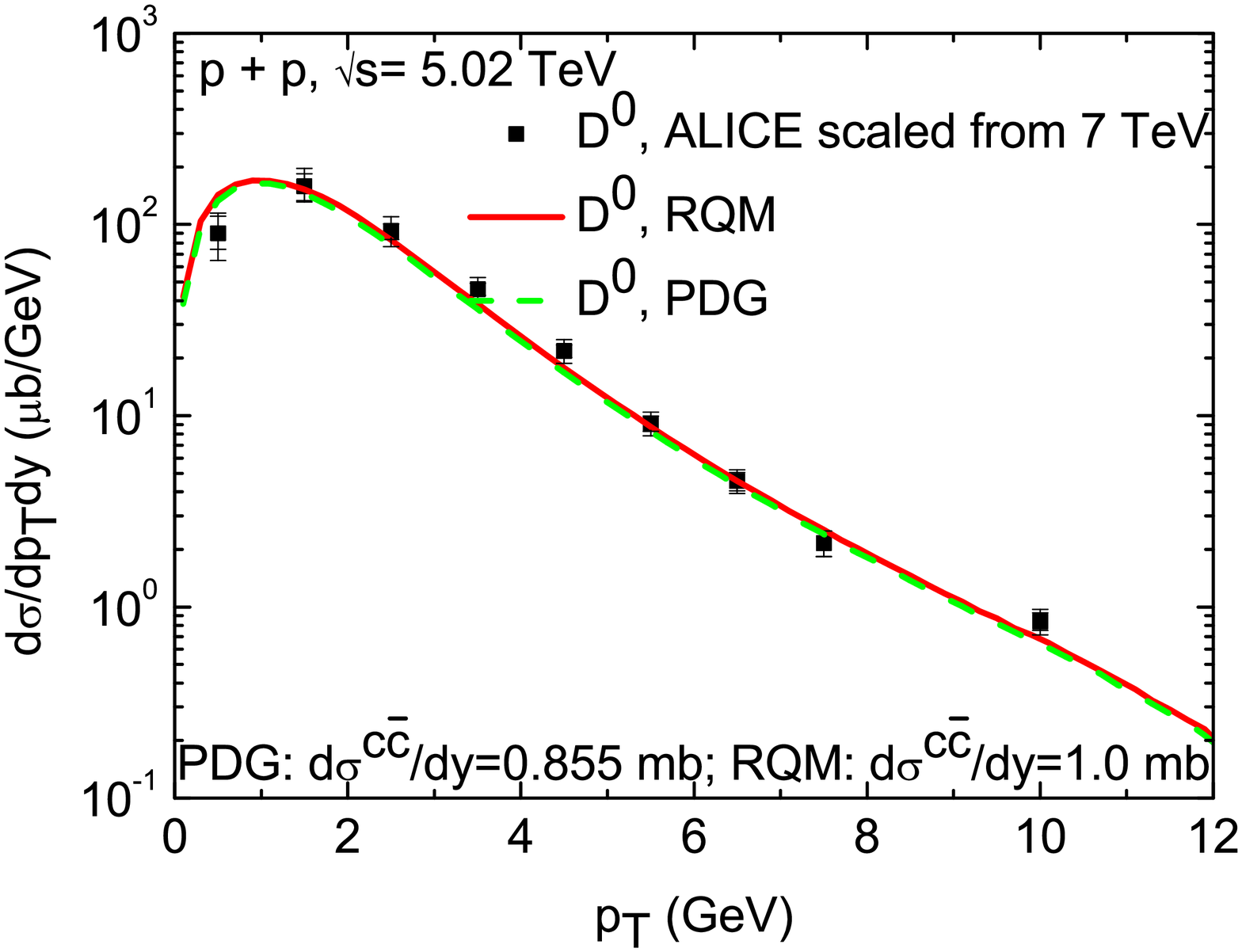}

\vspace{-0.75cm}

\includegraphics[width=1.0\columnwidth]{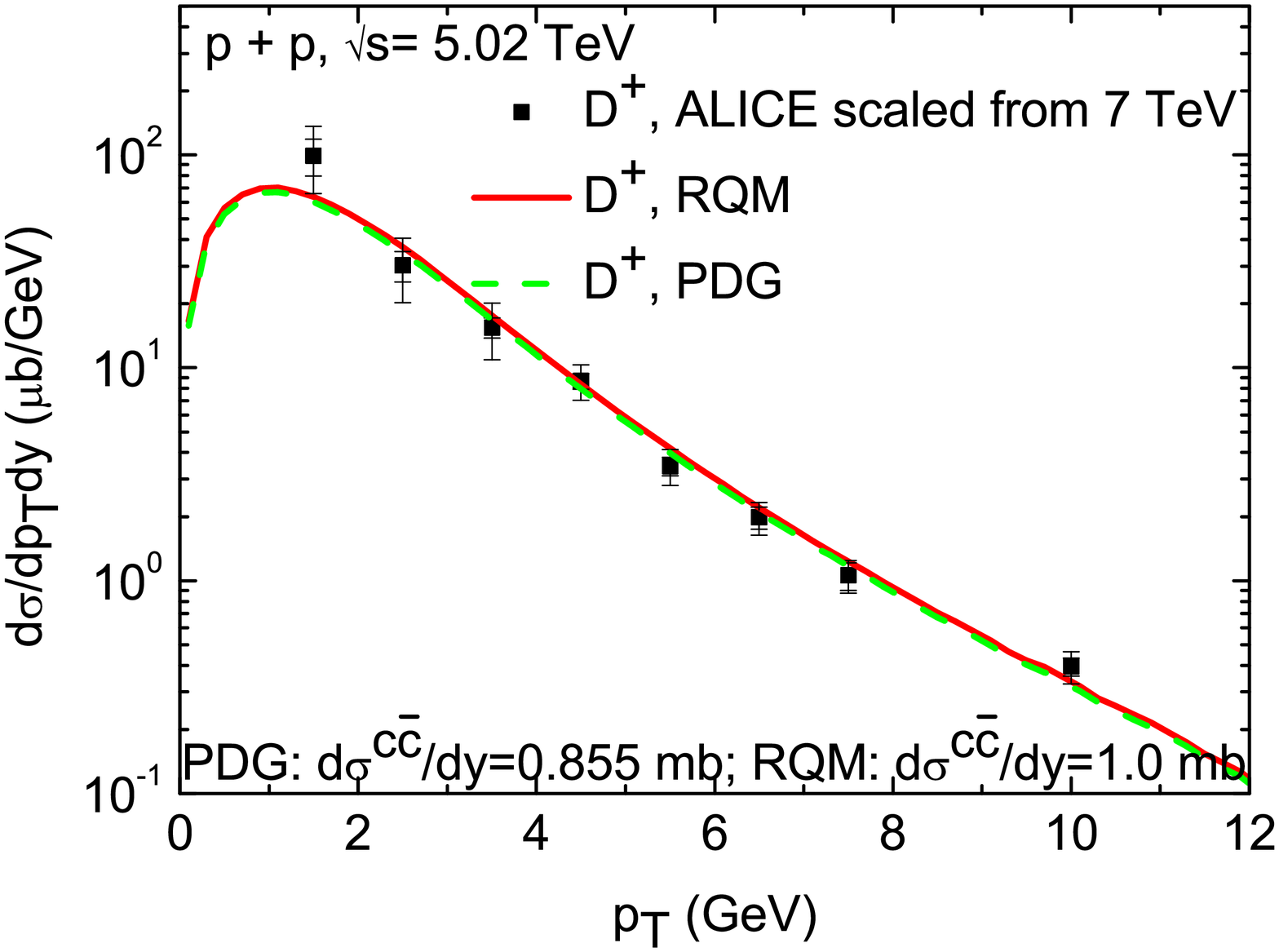}

\vspace{-0.75cm}

\includegraphics[width=1.0\columnwidth]{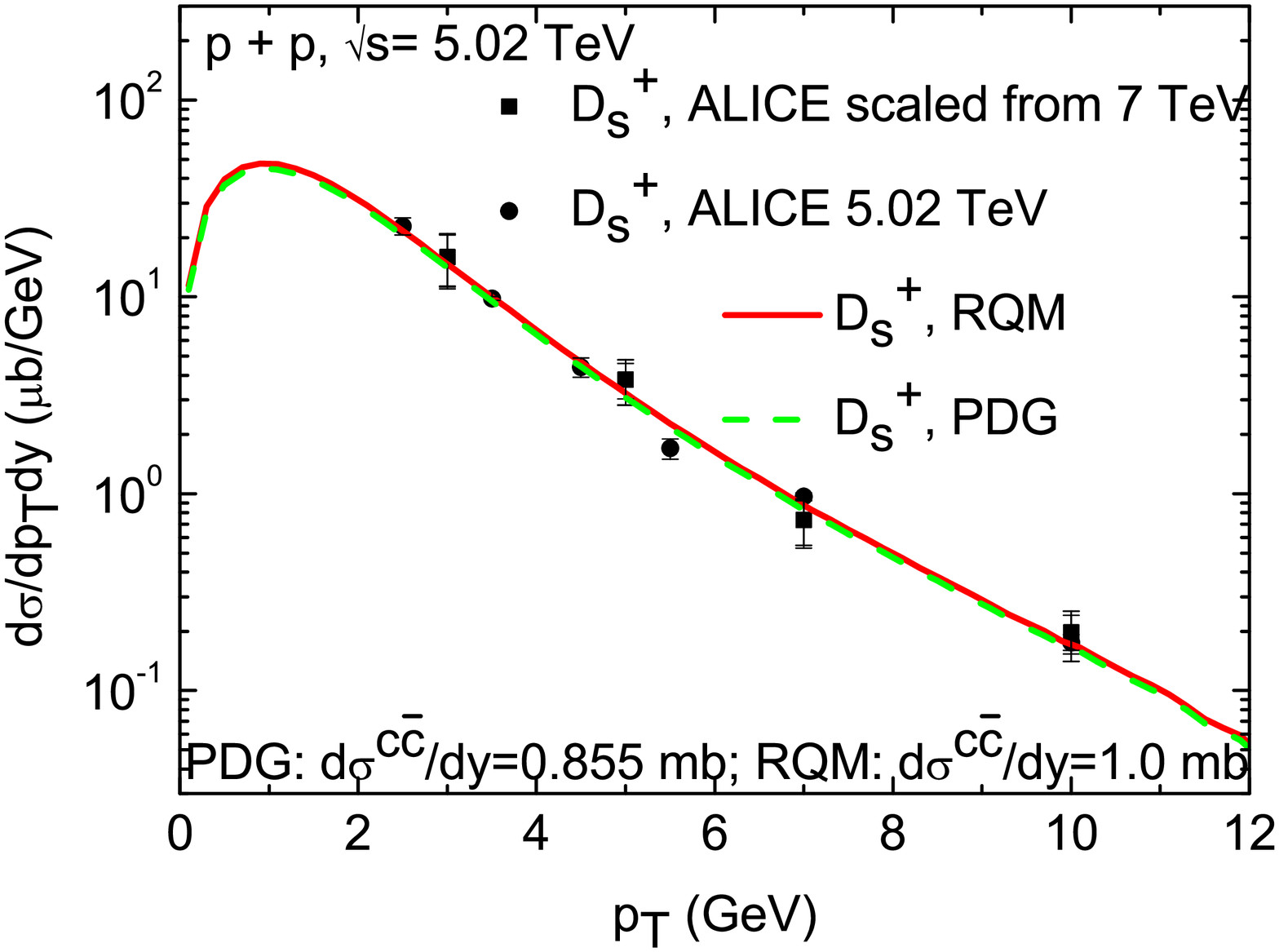}

\vspace{-0.75cm}

\includegraphics[width=1.0\columnwidth]{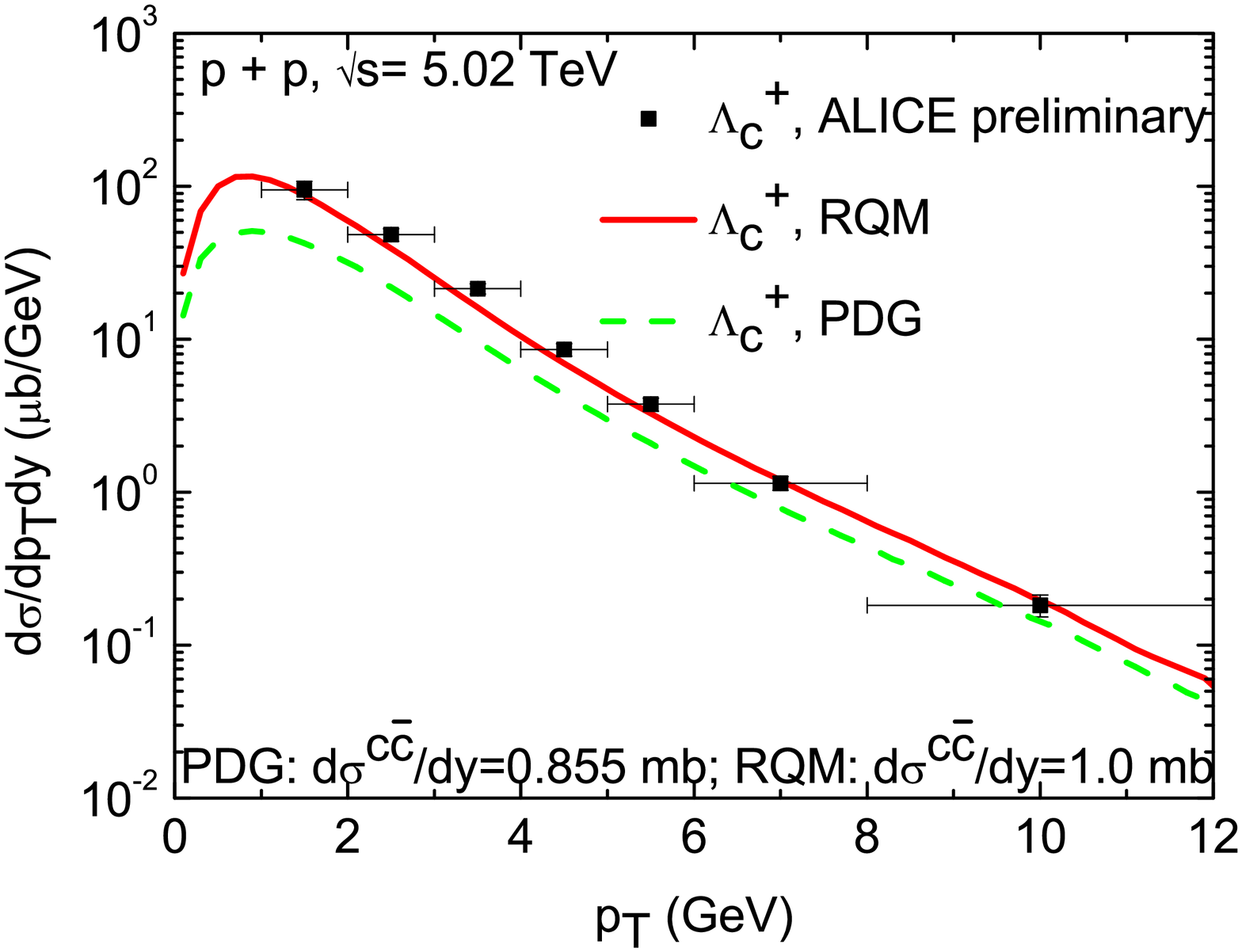}

\vspace{-0.4cm}

\caption{(Color online) Cross sections of $D^0$, $D^+$, $D_s^+$ and $\Lambda_c^+$ (including strong feeddowns) as
a function of transverse momentum in $\sqrt{s}=5.02$\,TeV $pp$ collisions. The PDG (dashed green lines) and RQM
scenario (solid red lines) with $T_{H}=170$\,MeV are compared to ALICE data at
mid-rapidity~\cite{Adam:2016ich,Acharya:2019mgn,Acharya:2017kfy}.
}
\label{fig_pT502}
\end{figure}
The results of the fits of the fragmentation and decay simulations in both PDG and RQM scenarios with $T_{H}=170$\,MeV
to the $p_T$-differential cross sections, $d\sigma/dp_Tdy$, for $D^0$, $D^+$, $D_s^+$ and $\Lambda_c^+$, as measured by
ALICE in $\sqrt{s}=5.02$\,TeV $pp$ collisions, are shown in Fig.~\ref{fig_pT502} together with the data~\cite{Adam:2016ich,Acharya:2019mgn,Acharya:2017kfy}.
The fitted total charm cross sections turn out to be $d\sigma^{c\bar{c}}/dy=0.855$\,mb and $1.0$\,mb
in these two scenarios, respectively. While the meson spectra can be well reproduced within the PDG scenario, the
$\Lambda_c^+$ spectrum exhibits a substantial deficiency in this scenario. Including the additional RQM baryons makes a
decisive difference and enables a good description of the $\Lambda_c^+$ spectrum measured by ALICE. Also note that the decay feeddown leads to an appreciable low-$p_T$ enhancement  over the PDG scenario that seems to be supported by the ALICE data.

\begin{figure} [!t]
\includegraphics[width=1.05\columnwidth]{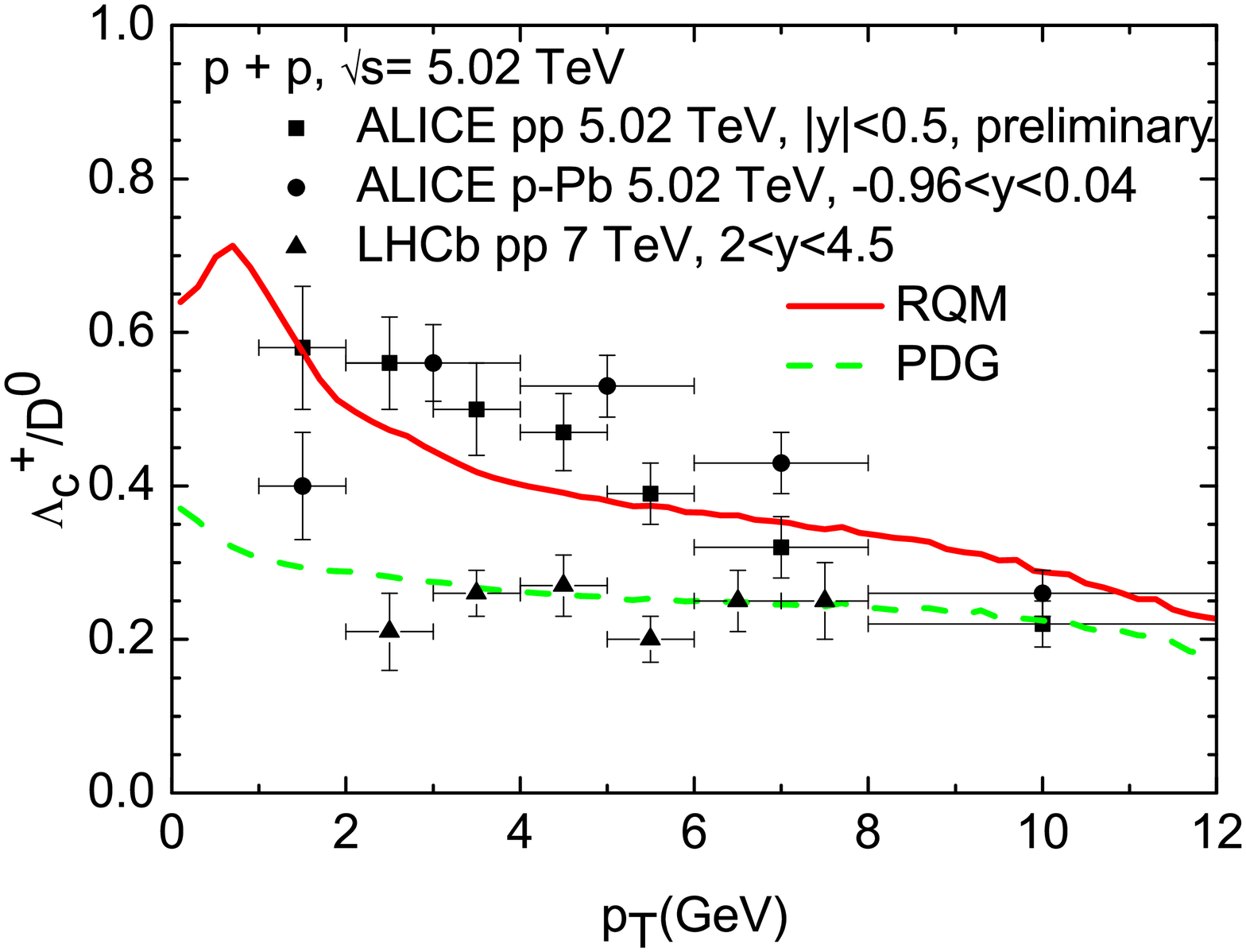}
\vspace{-0.5cm}
\caption{(Color online) The $\Lambda_c^+/D^0$ ratio following from our fits in the  PDG (dashed green line) and
RQM (solid red line) scenario at $T_{H}=170$\,MeV in $\sqrt{s}=5.02$\,TeV $pp$ collisions, compared to
ALICE~\cite{Acharya:2017kfy} and LHCb~\cite{Aaij:2013mga} data.
}
\label{fig_LcD502}
\end{figure}

Next, we turn to the  $\Lambda_c^+/D^0$ ratio following from our fit, shown in Fig.~\ref{fig_LcD502}. The ALICE
data~\cite{Acharya:2017kfy} at mid-rapidity confirm that  the RQM scenario is clearly favored, including the increasing trend toward
lower $p_T$ as referred to above. On the other hand, the LHCb data~\cite{Aaij:2013mga} at forward rapidity are better
reproduced by the PDG scenario.
Possible reasons for this may be the reduced particle multiplicity, \ie, the fewer production of light quarks and antiquarks at forward
rapidity, which limits the phase space available for charm-quark coalescence especially for more massive resonances, or a lower
hadronization temperature~\cite{Becattini:2007qr}.
Interestingly, the LHCb data in $p$-Pb collisions~\cite{Aaij:2018iyy} show an increase in this ratio, and possibly also a rising trend
toward central rapidity, while the ALICE $p$-Pb data at midrapidity are consistent with their $pp$ data~\cite{Acharya:2017kfy}, possibly exhibiting a slight hardening.

%%%%%%%%%%%%%%%%%%%%%%%%%%%%
\subsection{Predictions at  RHIC energy   \label{ssec_rhic}}
\begin{figure} [!t]
\includegraphics[width=1.05\columnwidth]{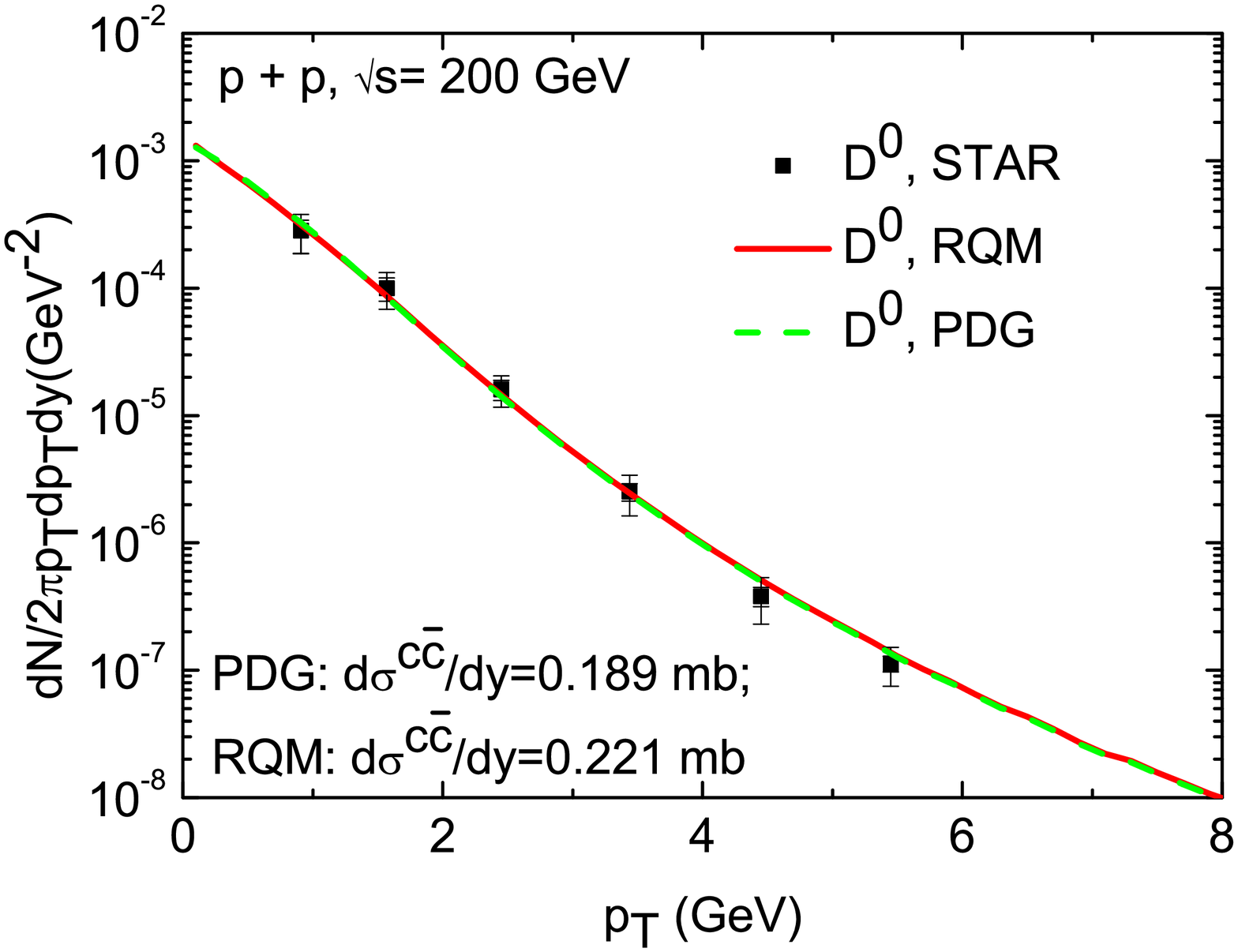}

\vspace{-0.6cm}

\includegraphics[width=1.05\columnwidth]{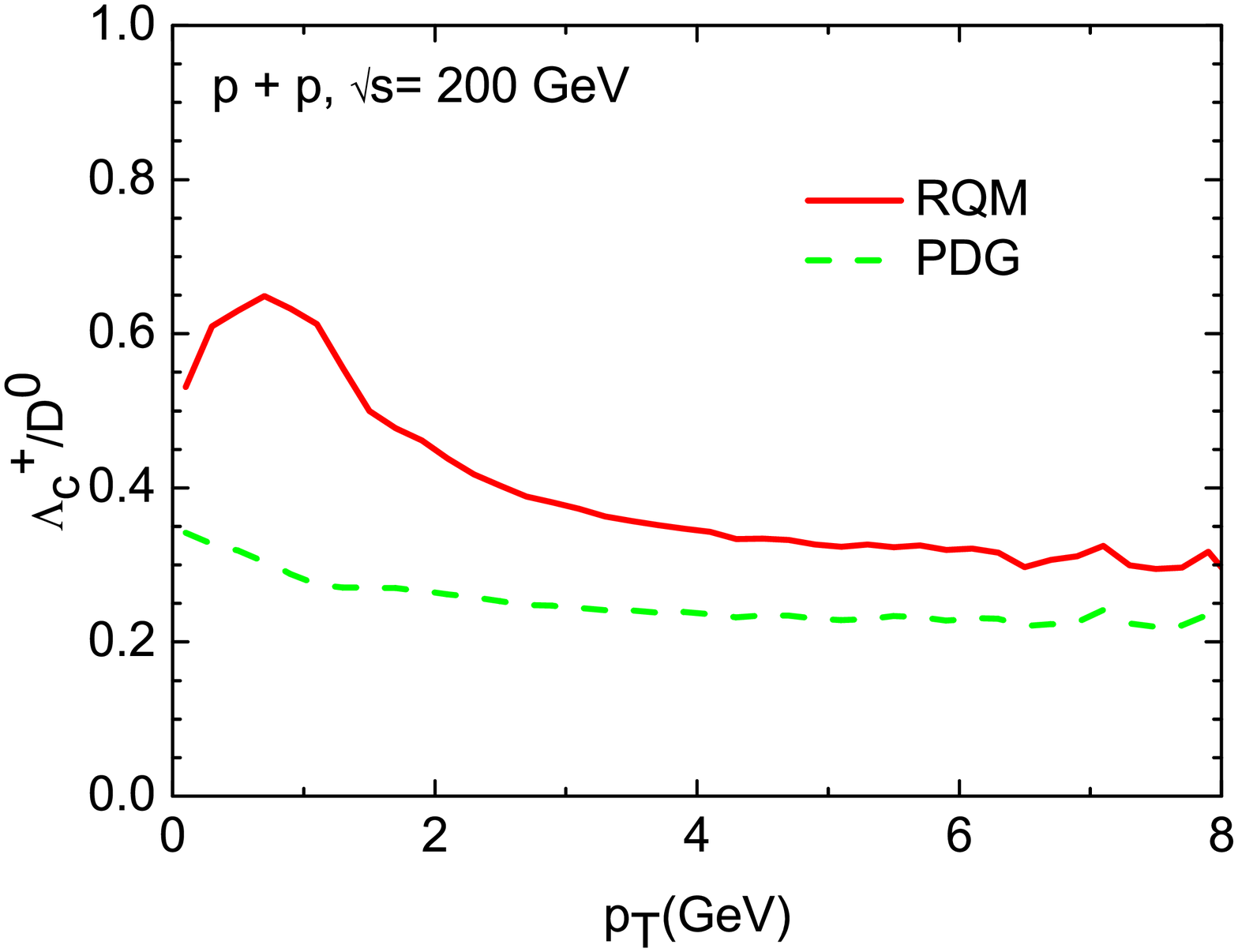}

\vspace{-0.4cm}

\caption{(Color online) The $p_T$ spectrum of $D^0$ 's (upper panel) and the $\Lambda_c^+/D^0$ ratio (lower panel) in
$\sqrt{s}=200$\,GeV $pp$ collisions from PDG and RQM scenarios at $T_H=170$\,MeV, together with STAR $D^0$ data in the
upper panel~\cite{Adamczyk:2012af,Adamczyk:2014uip}.
}
\label{fig_DLc-rhic}
\end{figure}

%%%%%%%%%%%%%%%%%%%%%%%%%%%
We repeat our fragmentation and decay simulation in the RHIC energy regime, for $\sqrt{s}=200$\,GeV $pp$ collisions, with
the same parameters for both PDG and RQM scenarios at $T_H=170$\,MeV. The only change is the underlying charm-quark $p_t$
spectrum which we again adopt from the FONLL framework, and the total charm input cross section. The thus obtained
$p_T$ spectrum for $D^0$'s is plotted in Fig.~\ref{fig_DLc-rhic} and shows good agreement with STAR
data~\cite{Adamczyk:2012af,Adamczyk:2014uip}.  The fitted charmed cross section turns out to be
$d\sigma^{c\bar{c}}/dy=0.221~(0.189)$\,mb in the RQM (PDG) scenario. The pertinent predictions for the
$\Lambda_c^+/D^0$ ratio are displayed in Fig.~\ref{fig_DLc-rhic}, showing very similar features as at LHC energies.

%%%%%%%%%%%%%%%%%%%%%%%%%%%%%%%%%%%%%%%%%%%%%%%%%%%%%%%%%%%%%%%%
\section{Summary
\label{sec_sum}}
%%%%%%%%%%%%%%%%%%%%%%%%%%%%%%%%%%%%%%%%%%%%%%%%%%%%%%%%%%%%%%%%
We have employed the statistical hadronization model to compute the hadro-chemistry of charm hadrons in $pp$ collisions
at collider energies. In particular, we have augmented the underlying charm-baryon spectrum by a relatively large number of states
as predicted by the relativistic quark model. A related issue is well known for the spectroscopy of  light and strange baryons, where
many more states are observed than in the charm sector. The need for additional charm-baryon  states is further
supported by lattice-QCD computations of the vacuum spectrum and of thermal charm susceptibilities in the vicinity of the transition temperature. Utilizing the RQM spectrum, we have found a marked increase of the $\Lambda_c/D^0$ ratio over the predictions
based on known states. As a result, the surprising enhancement of $\Lambda_c$ production as found by the ALICE collaboration in
$\sqrt{s}=7$\,TeV $pp$ collisions at midrapidity can be explained within the theoretical and experimental uncertainties (the smaller enhancement found at forward rapidities by the LHCb collaboration may hint at limitations of this picture). We have also computed pertinent $p_T$ spectra using the fragmentation function formalism but with the hadron ratios determined by the SHM. Also here
a good agreement with data has been found, including a low-$p_T$ enhancement for the $\Lambda_c$ which we atrribute to
feeddown from excited states.

Our findings suggest several directions of future work. The augmented SHM  can be tested by other charm hadrons (such as
 $\Sigma_c$ or $\Xi_c$) in $pp$ and $p$-A collision. It also has predictive power for the bottom
sector. Furthermore, we expect that our findings have important ramifications for the understanding of the charm (and bottom)
hadro-chemistry and kinetics in heavy-ion collisions. The intense rescattering of charm (and presumably also bottom) quarks in the hot
QCD medium, as reflected by the nuclear modification factor and large elliptic flow of $D$-mesons in Au-Au and Pb-Pb collisions
at RHIC and the LHC, stipulates the need for a controlled and universal equilibrium limit in transport calculations of the spectra
and yields of charm hadrons at low and intermediate $p_T$~\cite{Rapp:2019bxp}. We believe that our analysis presented here
provides a significant and well-motivated improvement in this direction, not only for understanding $\Lambda_c$
production~\cite{Acharya:2018ckj,Zhou:2017ikn}, but also for current and future measurements of a much richer set of
charm ($D_s$, $\Sigma_c$, ...) and bottom ($B_s$, $\Lambda_b$, $\Sigma_b$, ...) hadrons.
\\

{\bf Acknowledgments:}
This work was supported by the NSFC under grant 11675079 (MH), and by the U.S.~National Science Foundation (NSF) under grant no.~PHY-1614484 (RR).

\end{document}